\documentclass{sigcomm-alternate}

\usepackage[pass]{geometry}

\PassOptionsToPackage{hyphens}{url}\usepackage{hyperref}
\usepackage{color}
\usepackage{booktabs}
\usepackage{graphicx}
\usepackage{multirow}
\usepackage{subfig}
\usepackage{pifont}
\usepackage{xspace}
\usepackage{cite}
\usepackage{times}
\usepackage{microtype}
\usepackage{balance}
\newsubfloat{subfigure}
\hypersetup{pdfborder=0 0 0,
  colorlinks=true,
  citecolor=darkblue,
  linkcolor=darkblue,
  urlcolor=darkblue
}
\newcommand{\xref}[1]{Section~\ref{#1}}
\newcommand{\cref}[1]{Chapter~\ref{#1}}

\newcommand{\fref}[1]{Fig.~\ref{#1}}
\newcommand{\tref}[1]{Table~\ref{#1}}

\newcommand{\first}{\emph{(i)}~}
\newcommand{\second}{\emph{(ii)}~}
\newcommand{\third}{\emph{(iii)}~}
\newcommand{\fourth}{\emph{(iv)}~}
\newcommand{\fifth}{\emph{(v)}~}

\newcommand{\ie}{i.e., \@}
\newcommand{\eg}{e.g., \@}

\newcommand{\Eg}{For example, \@}
\newcommand{\cf}{cf. \@}

\newcommand{\etal}{et~al.\xspace}

\newcommand{\perc}{\,\%\xspace}

\newcommand{\checked}{\ding{51}\xspace}

\newenvironment{smallenumerate}{%
\begin{enumerate}%
\setlength{\parsep}{0.3ex}%
\setlength{\leftmargin}{0.1pt}%
\setlength{\listparindent}{-8ex}%
\setlength{\itemsep}{0.2ex}%
\setlength{\topsep}{0ex}%
\setlength{\parskip}{0ex}%
\setlength{\partopsep}{0ex}%
}{%
\end{enumerate}
}%

\definecolor{darkgreen}{rgb}{0,0.5,0}
\definecolor{brown}{rgb}{0.7,0.3,0}
\definecolor{darkblue}{rgb}{0,0,0.5}

\newcounter{fn1}
\newcounter{fn2}
\newcounter{fn3}
\newcounter{fn4}
\newcounter{fn5}
\begin{document}
%\frontmatter          % for the preliminaries
%\pagestyle{headings}  % switches on printing of running heads
%\mainmatter              % start of the contributions
%
\DeclareFixedFont{\auacc}{OT1}{phv}{m}{n}{12}
\title{A Comparative Look into Public IXP Datasets}
\numberofauthors{1}
\author{
\begin{tabular*}{\textwidth}%
{@{\extracolsep{\fill}}ccccc}
Rowan Kl{\auacc\"o}ti & Bernhard Ager & Vasileios Kotronis & George Nomikos & Xenofontas Dimitropoulos\\
\multicolumn{3}{c}{\affaddr{ETH Zurich, Switzerland}} & \multicolumn{2}{c}{\affaddr{FORTH, Greece}}\\
\multicolumn{3}{c}{\email{\{rkloeti,bager,vkotroni\}@tik.ee.ethz.ch}} & \multicolumn{2}{c}{\email{\{gnomikos,fontas\}@ics.forth.gr}}\\
\end{tabular*}
}

%\author{Rowan Kl\"oti\inst{1} \and Bernhard Ager\inst{1} \and \\ Vasileios Kotronis\inst{1}
%\and Xenofontas Dimitropoulos\inst{1,2}}

%\titlerunning{A Comparative Look at Public IXP Directories: Partially Consistent}
%\authorrunning{R. Kl\"oti, B. Ager, V. Kotoronis, X. Dimitropoulos}
%% \institute{ETH Zurich, Zurich, Switzerland,\\
%% \email{$\lbrace$rkloeti,vkotroni,bager$\rbrace$@tik.ee.ethz.ch}
%% \and
%% Foundation of Research and Technology Hellas (FORTH), Greece\\
%% \email{fontas@ics.forth.gr}}

\maketitle              % typeset the title of the contribution

\begin{abstract}

Internet eXchange Points (IXPs) are core components of the Internet infrastructure
where Internet Service Providers (ISPs) meet and exchange traffic. During the last few
years, the number and size of IXPs have increased rapidly, driving the flattening and
shortening of Internet paths. However, understanding the present status of
the IXP ecosystem and its potential role in shaping the future Internet requires 
rigorous data about IXPs, their presence, status, participants, etc. 
 %Internet Exchange Points (IXPs) are becoming increasingly central entities of
%the Internet ecosystem due to the flattening of the Internet topology and have
%recently attracted vigorous research interest. 
%However, previous work on IXP
%data is limited to a single data source, namely PeeringDB, and is therefore
%incomplete. 
In this work, we do the first cross-comparison of three well-known publicly
available IXP databases, namely of PeeringDB, Euro-IX, and PCH. A key
challenge we address is linking IXP identifiers across databases maintained
by different organizations. We find different \emph{AS-centric} versus \emph{IXP-centric}
views provided by the databases as a result of their data collection approaches. In
addition, we highlight differences and similarities w.r.t. IXP participants, 
geographical coverage, and co-location facilities. As a side-product of our linkage
heuristics, we make publicly available the union of the three databases, which
includes 40.2\perc more IXPs and 66.3\perc more IXP participants than the
commonly-used PeeringDB. We also publish our analysis code to foster
reproducibility of our experiments and shed preliminary insights into the
accuracy of the union dataset.

\keywords{Internet Exchange Points, Peering, Euro-IX, PeeringDB, PCH}
\end{abstract}
\section{Introduction}\label{sec:intro}

A large part of the interconnection between Autonomous Systems (ASes) in the Internet is realized via
\emph{Internet eXchange Points} (IXPs), giving them a major role in the evolution and
performance of the Internet. Notably, researchers have recently found that
\first the Internet topology is flattening due to IXP-traversing paths which bypass the
classic transit hierarchy~\cite{gill2008flattening,FlatInternet,IXPStructure,Labovitz:2010:IIT:2043164.1851194},
\second more peerings exist in a single large IXP than in previous sets of measurements for the
entire Internet~\cite{Ager:2012:ALE:2377677.2377714}, and \third end-to-end delays
and path lengths over IXPs are becoming shorter~\cite{IXPsInternetDelays}. Furthermore,
IXPs have been proposed as cradles for hosting new technologies, such as Software
Defined eXchanges (SDX)~\cite{SDX-SIGCOMM}.

%%\todo{If we cite SDX we should also cite CXP here.}

%Previous work
%has explored primarily information from PeeringDB~\cite{peeringdb-url}, a user-maintained IXP directory for AS
%administrators, to find and verify IXP member ASes. However, there are several additional
%sources of systematic information on IXPs: \emph{The European Internet Exchange Association}
%(Euro-IX) maintains a list of IXPs~\cite{euix-url}, while \emph{Packet Clearing House} (PCH)
%maintains a directory~\cite{pch-ixp-dir}, which also includes many historical IXPs.
%These datasets are collected to varying degrees of detail either manually or automatically
%by the publisher, or through self-reporting by the IXPs and their participants.
%We aim to find out how these datasets compare with each other in terms of consistency and completeness.

However, the merits and artifacts of the available IXP data have not been thoroughly
researched yet. This is in sharp contrast with extensive research on mapping the
interconnections between ASes using data from various sources, like RouteViews~\cite{routeviews},
for more than a decade. A commonly-used source of IXP data in scientific studies,
\eg~\cite{IXPMap,peeringdb-routing-ecosystem,Chatzis:2013:MIM:2541468.2541473},
is PeeringDB~\cite{peeringdb-url}. However, in addition to PeeringDB, two other publishers
maintain public databases about the global IXP ecosystem, namely the \emph{European Internet
Exchange Association} (Euro-IX)~\cite{euix-url} and \emph{Packet Clearing House} (PCH)~\cite{pch-ixp-dir}.
These datasets are contributed and kept up-to-date by different stakeholders, \eg by the publisher,
or through self-reporting by the IXPs and their participants.

In this work, we do the first cross-comparison of the IXP data provided by PeeringDB, Euro-IX,
and PCH. We compare in depth several attributes, like IXPs' locations, facilities and 
participant information. We highlight the similarity of the available data, complementary
information, and data discrepancies. We analyze in total data from about 499, 490, and 687 IXPs
in PeeringDB, Euro-IX, and PCH, respectively. To compare the data, we introduce heuristics
to link identical IXPs across the three datasets. We find an \emph{IXP-centric} view provided
by Euro-IX vs. an \emph{AS-centric} view provided by PeeringDB, reflecting differences
in their often volunteer-based data collection approaches.

Besides, we make the linked datasets and our analysis code publicly available~\cite{mappings}
to support reproducibility of our experiments and related research efforts. Experiments where
this data can be useful include, but are not limited to, \first discovering new peer-to-peer links based on membership data 
and peering policy so as to augment the Internet 
topology view, \eg for modeling the effect of augmented routing protocols~\cite{bgp-security-in-partial-deployment}, 
\second investigating the peering ecosystem from a geographical perspective per continent or country,
\third tracking the historic evolution of IXPs and their features, \fourth pinpointing the big players in a peering setup,
and \fifth working with new topological paradigms such as IXP multi-graphs~\cite{kotronis2429control} in the
context of new service provisioning. Compared to using solely PeeringDB, the union of the linked datasets includes data
for 40.2\perc more active IXPs and 66.3\perc more IXP participants. 

Finally, we perform a preliminary analysis of the accuracy of the linked datasets and
find that even the combined dataset is only 75\perc complete when comparing with
information from BGP route collectors, indicating the need for
further research in this context. Partial verification using data available on IXP websites shows
more promising results in terms of accuracy, both for the biggest IXPs and for
IXPs that are randomly selected from the combined pool of available IXPs.
We would like to note though that the three IXP datasets are collected based on voluntary effort 
and as such, no formal guarantees about completeness, accuracy or freshness can generally be given.

The rest of this paper is structured as follows. We first discuss differences and similarities in particular w.r.t.
the collection methodologies of the PeeringDB, Euro-IX, and PCH datasets in~\xref{sec:data_sources}.
Then, we introduce our heuristics to link IXPs across datasets in~\xref{sec:matching}.  We compare the
IXP status, location, and facility information in~\xref{sec:analysis} and the IXP participant information
in~\xref{sec:ixp-participants}. We discuss and evaluate the accuracy of the datasets in~\xref{sec:accuracy}.
Finally, \xref{sec:conclusions} concludes our paper and points to future directions.

\section{Data Sources}\label{sec:data_sources}

We analyze and cross-compare the three most extensive publicly
available IXP datasets, which are provided by PeeringDB~\cite{peeringdb-url},
Euro-IX~\cite{euix-url}, and PCH~\cite{pch-ixp-dir}. The datasets
inform primarily about IXPs and their participants in varying levels
of detail. In \tref{tab:attributes} we compare the types of information
and their level of availability in each of the datasets. 
Importantly, naming and location information is contained in all datasets,
enabling us to identify and link identical IXPs in~\xref{sec:matching}. 
We built custom web crawlers and parsers, which we make publicly
available~\cite{mappings}. A crawl typically takes between 10 and
30 minutes, depending on the dataset. We acquired all datasets on September 19, 2014.
In the remainder of this section, we discuss intrinsic characteristics of each
dataset, shedding light on the underlying methodology used by the three data
providers to collect and maintain the data.

\subsection{PeeringDB}\label{sec:peeringdb}
PeeringDB~\cite{peeringdb-url} is a worldwide database that aims to serve ISPs
which wish to participate in the IXP peering ecosystem. The data available
consists of 499 IXPs, their facilities and their participants (\ie peering ASes).
%This contrasts with Euro-IX, which provides no information about IXPs'
%facilities, except for the approximate geographical location. \fontas{This has been already discussed above.}
PeeringDB has detailed information about all registered IXPs, unlike
Euro-IX which only has detailed information about its affiliate IXPs,
while data on non-affiliate IXPs is limited to name, location and status.
Moreover, PeeringDB provides detailed information about individual
participants, \ie ASes that peer at IXPs. The data is self-reported by both IXPs and participants.

%The correctness and completeness of the PeeringDB data 
%has been investigated in previous work~\cite{peeringDB-accuracy,peeringdb-routing-ecosystem}.
%Snijders~\cite{peeringDB-accuracy} found that 75\% of the IXP memberships 
%in 256 \texttt{show bgp sum} dumps provided by network operators were listed in PeeringDB.
%Similarly, Lodhi \etal~\cite{peeringdb-routing-ecosystem} found a median of 80\perc
%completeness in the PeeringDB members of the 20 largest IXPs, with significant
%differences though between regions; \eg the completeness was only 25\perc for the
%Moscow IX. Both studies report that although the PeeringDB data is not complete,
%it is generally accurate~\cite{peeringDB-accuracy,peeringdb-routing-ecosystem}.

%The data provided by PeeringDB has been investigated in other research
%efforts and has been reported to be mostly accurate~\cite{peeringDB-accuracy,peeringdb-routing-ecosystem}
%in terms of network-specific properties; its accuracy regarding IXP-specific properties is not 
%fully verified, and completeness has not been checked~\cite{peeringdb-routing-ecosystem}.

%As such, it seems less likely that a bias towards certain IXPs exists,
%as all IXPs have an incentive to keep their data accurate and
%up-to-date, with PeeringDB being widely used amongst prospective IXP
%members.
%The members themselves are responsible
%for keeping the available information up-to-date.

\subsection{Euro-IX}\label{sec:eu_ix}
Our second dataset is a list of 490 IXPs provided by the European
Internet Exchange Association (Euro-IX)~\cite{euix-url}. Its
membership consists mostly of European IXPs, which are typically run
as cooperative non-profit entities, in contrast to North American
Internet exchanges, which are often run as for-profit businesses.
Accordingly, European Internet exchanges are generally transparent
about peering arrangements. 
Some of the largest IXPs are located in Europe.
Euro-IX supplies information both for affiliated and
non-affiliated IXPs. According to the official Euro-IX website~\cite{euix-url},
``the database information is a combination of both affiliated and 
non-affiliated IXP content. While the affiliated IXP content is 
highly accurate, the non-affiliated IXP content is updated on a 
best effort basis and is nonetheless considered to be quite accurate''.
From direct communication with Euro-IX staff,
we know that the information is generally provided by the IXPs
themselves. About two thirds of the IXPs represented have an account
to keep their data up-to-date by self-reporting, while 62 of these IXPs
(approximately 14\perc) have automated the update procedure,
which helps improve data completeness and accuracy. Euro-IX provides
a website URL and a contact email for all IXPs and participants, 
\ie the ASes which connect to an IXP, for 285 of the IXPs. 
For a subset of IXPs (we assume these are the ones which are
registered members of Euro-IX), more detailed information is
available (c.f. \tref{tab:attributes}). For IXP participants there is
limited information, including AS numbers (ASN), name, update
time-stamp, IPv6 support capability, and sometimes a URL.

Euro-IX does not provide details about IXPs' individual co-location
facilities. However, location information at the city level and, for
most IXPs, geographical coordinates are available. Since IXPs can be
distributed over several co-location facilities,
these location values may not accurately reflect the physical IXP location. 
For instance, \emph{CyrusOne} is a distributed (likely not
Euro-IX affiliated) IXP in Arizona and Texas with points of presence in Austin, Dallas, Houston, Phoenix
and San Antonio, but appears in the Euro-IX database only at Carrollton, a suburb 
of Houston, where its corporate headquarters are located. In addition, 
Euro-IX does not provide information about IP address prefixes assigned
to IXPs, which could potentially be used for linking IXPs across databases.

\newcommand{\always}{\checked}
\newcommand{\yes}{\checked}
\newcommand{\mostly}{$+$}
\newcommand{\usually}{$\circ$}
\newcommand{\sometimes}{$\circ$}
\newcommand{\no}{}
\newcommand{\rl}[1]{\rotatebox{90}{#1}}

\begin{table*}[t]
  %%% For adding footnotes, it's probably best to try the threeparttable package
  \centering
  \scriptsize
  \tabcolsep1.7pt
  \begin{tabular}{l*{32}{c}}
    \toprule
      & \multicolumn{18}{c}{IXP} & \multicolumn{12}{c}{Members} & \\
              %\cmidrule(r{0.25em}){1-1}
              \cmidrule(r{0.25em}l{0.25em}){2-19}\cmidrule(l{0.25em}){20-32}
    Data set         & \rl{Country and city} & \rl{Continent} & \rl{Coordinates} & \rl{Long Name} & \rl{Common Name} & \rl{Status (active)} & \rl{Media Type}\rl{(Ethernet, etc)} & \rl{Protocols}\rl{supported} & \rl{Website} & \rl{Contact}\rl{information} & \rl{Costs} & \rl{Establishment date} & \rl{Membership}\rl{requirements} & \rl{AS Number} & \rl{Network internals} & \rl{Associated members} & \rl{\# facilities} & \rl{Detailed facility info} & \rl{Organization} & \rl{ASN} & \rl{IP address (at IXP)} & \rl{Company Name} & \rl{Company Website} & \rl{Protocols supported} & \rl{Date Last Updated} & \rl{URLs} & \rl{Network details} & \rl{Policy information} & \rl{Approx \# prefixes} & \rl{TXT Record} & \rl{Network status} \\
              \midrule
    Euro-IX   & \always & \no & \mostly & \usually & \always & \always & \sometimes & \yes & \mostly & \yes & \yes & \sometimes & \sometimes & \sometimes & \sometimes & \always & \yes & \no & \sometimes & \always & \no & \always & \mostly & \always & \no & \no & \no  & \no & \no & \no & \no \\
    PeeringDB & \always & \always & \no & \usually & \always & \no & \always & \yes & \yes & \yes & \no & \no & \no & \sometimes & \yes & \always & \always & \yes & \no & \always & \mostly & \always & \mostly & \mostly & \mostly & \mostly & \mostly & \mostly & \mostly & \no & \no \\
    PCH       & \mostly & \no & \no & \mostly & \mostly & \mostly & \mostly & \no & \yes & \no & \no & \sometimes & \no & \no & \sometimes & \sometimes & \no & \no & \sometimes & \sometimes & \mostly & \sometimes & \no & \no & \no & \no & \no & \no & \sometimes & \sometimes & \yes \\
    \bottomrule
  \end{tabular}
  \caption{Comparison of information available from the Euro-IX, PeeringDB, and PCH datasets. Available = \yes, mostly available = \mostly, sometimes available = \sometimes.}
  \label{tab:attributes}
\end{table*}

\subsection{Packet Clearing House}\label{sec:pch}
The Packet Clearing House (PCH) 
is a non-profit research institute
concerning itself with Internet routing and traffic exchange, among
other areas pertaining to Internet operation and economics.
%such as DNS anycast services. 
PCH provides an extensive directory of 687 IXPs~\cite{pch-ixp-dir}, including many historical ones.
Indeed, Chatzis \etal~\cite{Chatzis:2013:MIM:2541468.2541473} claim that PCH never
removes IXPs from the listing, and marks them defunct only after sufficient verification.
According to direct communication with PCH staff, 70\perc of the IXPs listed are compiled by PCH
staff, 25\perc are contributed by the Internet community and some
5\perc are added by the IXP operators themselves. PCH peers at
many IXPs itself; the BGP information PCH obtains over these
peerings is then used to derive participant lists. PCH also compiles
traffic data from MRTG files (for 24 IXPs); the other data sources do not
have automatic traffic information.
%% PCH peers at many many more IXPs than only 24. --Bernhard
%This means that
%such lists are generally only available for IXPs where PCH is
%present through peering; consequently, participant information is
%not available for most IXPs. 
For 190 subnets (corresponding to nearly as many IXPs)
participant data is entered manually.
PCH reports on a per-port basis, not a per-participant basis.
As such, an ASN can appear multiple times as a member of an IXP.
There are also numerous instances of participant entries containing peering IP
addresses but no ASNs. We only consider entries with ASNs, as we have no other
consistent basis for matching the participants across datasets.

\subsection{Data Artifacts}\label{sec:analysis-consistency}
During our data pre-processing and analysis, we observed several artifacts
(some quite time consuming) in the datasets, which we report here to
simplify future researchers' work.

PeeringDB has two sources of information
on connectivity between IXPs and ASes. For each IXP, there is a list of
participants, including ASNs. However, for every participant, there is
also a list of IXPs. These do not necessarily coincide. A quarter of
IXPs present in the PeeringDB dataset have differences between the two
sources of information, with more ASNs being listed in the
participants' IXP list. This is a consequence of the fact that some
participants advertise more than one ASN. The difference in terms of
number of participants is 5.7\perc on average, although typically no
more than a handful of entries. Only 0.5\perc of ASNs are responsible
for this difference. In general, using the latter data source
(participants' IXPs) is preferable due to a slightly higher completeness.

The Euro-IX dataset has 20 IXPs whose participants consist partially,
and nine whose participants consist entirely of the reserved ASN
``0''. In these cases, the administrator has apparently neglected to
enter an ASN. These participants contribute about 2\perc of the
participant entries present, and there are no other duplicate entries.

We also note that PCH has 39 IXPs which have multiple participant entries with the same ASN, with 237 ASNs
duplicated in total. Many others have no associated ASN reported at all. As noted in~\xref{sec:pch},
this is a result of the port-based reporting used by PCH.

\section{Linking IXPs across datasets}\label{sec:matching}

In this section we describe our methodology for identifying and
linking identical IXPs in different datasets as well as other 
pre-processing steps that were necessary to sanitize the
data. We use the term \emph{mapping} to refer to
identical IXPs that have been linked in two datasets.
The key challenge is that
IXPs lack consistent identifiers across the datasets. There are several
cases of IXPs sharing the same name when they are separate entities,
and many cases of identical IXPs being represented by different names in
the three datasets. An example is `SIX'---a name that occurs with minor
variations 5 times in PeeringDB (\ie SIX, S-IX, SIX.SK, SIX SI, SIX NO for Seattle-, Stuttgart-, Slovak-, Slovenian-, Stavanger- IXP respectively). In Euro-IX, there are only three 
variations of `SIX', two of which do not directly match the ones in PeeringDB,
and at least two different IXPs in Euro-IX share the exact name `SIX'.
In addition, for various reasons (\ie geographically distributed IXPs), some IXPs exist
as single entities in one dataset and as multiple entities in the other. 

%%, IXPs with multiple peering networks

%In an extreme case we found an identifier that was referring to different IXPs in two datasets.  

%\discusstodo{minimizing both false positives and false negatives is
%  not possible. moreover, we are performing semi-automated mapping -
%  it's irrelevant.}  While we wish to minimise both \emph{false
%  positives} (\ie IXP entries which are erroneously mapped together
%but do not correspond to the same entities) and \emph{false negatives}
%(\ie IXPs which are not mapped together despite the fact that they are
%the same entities), we avoid mapping entities when we cannot be sure
%of their identity. 

Due to the large number of IXPs in each dataset, linking all IXPs manually is
very tedious and time consuming. Unfortunately, a fully automated
approach is not desirable, either, as human expertise is necessary
to validate possibly ambiguous mappings. 
For these reasons, we use a hybrid approach, in which we first automatically
produce candidate mappings based on custom heuristics and then we
manually verify which candidates actually correspond to the same IXP. 
Our heuristics to generate candidates for mapping exploit IXP naming
and location information and are inclusive in their design. In other words
we are conservative in ruling out possible mappings,
at the cost of additional manual validation effort.

%As a first approach, we attempt to construct a mapping based on
%common participant information such as AS numbers. However, this
%turns out to be insufficient in practice due to inconsistent
%reing of participants (\cf \xref{sec:analysis-participants}). Our
%second approach, a semi-automatic mapping utilizing naming and
%location information, proves to be more fruitful. 

%Since all matches are
%manually verified, we expect no false positives other than human
%errors in the manual verification. In addition, our approach to generate
%candidates for mapping is inclusive. We therefore expect few
%false negatives only in cases for which we have insufficient or ambiguous
%information.

During our analysis
we found that IXPs are sometimes presented at different granularity in
the different datasets, \eg at a facility level in one dataset and as
a whole in another. Thus we first merge such sibling IXP records
into single entities using the same overall approach as with linking IXPs
across datasets. We produce mapping candidates for
IXPs that share the same name and location. 
We explored several schemes for transforming names in order
to get good mapping candidates between the different datasets. We apply
these name transforming schemes one-by-one, on the original name.  After
each step, we manually check the produced mappings and remove
successfully mapped IXPs from the working datasets. 
All datasets provide name aliases, which we also take into
consideration. Moreover, differences
in the location naming convention require additional pre-processing.

%For manual
%verification we utilize all available information for an IXP.

Overall, we first merge 26 sibling IXP records into 7 IXPs for a total of
471 IXPs in the Euro-IX dataset, 30 siblings into 12 IXPs
for a total of 480 IXPs in the PeeringDB dataset, and 47 siblings
into 18 IXPs for a total of 657 IXPs in the PCH dataset. We
then use the following heuristics to produce candidates 
(with the results for Euro-IX/PeeringDB,
Euro-IX/PCH, PeeringDB/PCH being respectively reported next to each variant):
\begin{smallenumerate}
\item Directly identical names (214 / 184 / 162 mappings)
\item Converting to lower case (16 / 21 / 26 new mappings).
\item Truncating the name at the second word boundary (2 / 15 / 3 new mappings).
\item Truncating the name at the first word boundary (67 / 101 / 76 new mappings).
\item Removing non-word characters (4 / 8 / 8 new mappings).
\item Various combinations of these, and manual matching (the remaining mappings).
\end{smallenumerate}
We also explored
heuristics based on common IXP member information such as ASNs.
However, this turned out to be insufficient in practice due to incomplete
reporting of IXP member ASNs (\cf \xref{sec:analysis-membership}). 
Another possible attribute that could be explored for linking is assigned 
IXPs' IP address prefixes. This data is provided by PeeringDB and PCH,
but not by Euro-IX. We therefore did not consider it. 

In total we find 380, 379 and 344 mappings, respectively.
Table~\ref{tab:matches-by-dataset} shows
the size of the intersection (the IXPs that match based on the
previous process) and the union (all IXPs) of the
datasets, as well as the Jaccard index and overlap index between two
sets $A$ and $B$ defined as:
\begin{math} J(A,B) = \frac{\vert A \cap B \vert}{\vert A \cup B \vert} \end{math} and
\begin{math} O(A,B) = \frac{\vert A \cap B \vert}{\min(\vert A
    \vert,\vert B \vert)} \end{math}. Intuitively, the Jaccard index
indicates the similarity between sets, while the overlap index
indicates the degree to which the smaller set is a subset of the
larger. We include both in order to indicate the extent to which the
difference is simply the result of one dataset being more complete
than the other, rather than the datasets being partially orthogonal.
For comparing all three sets we use straight-forward extensions of
the Jaccard and overlap indices, using all three sets as parameters.
%%instead of two to the intersection, union and min functions. 
All mappings have been manually verified and our approach to generate
candidates for mapping is inclusive as explained beforehand. We therefore do not expect false
mappings, but we could have missed few mappings in cases we had
insufficient or ambiguous information.

%\begin{table*}[t]
%\centering
%\scriptsize
%\begin{tabular}{ccccrrrrrrrr}
%\toprule
%& & & \multicolumn{4}{c}{All IXPs} & \multicolumn{4}{c}{Only active IXPs}\\
%\cmidrule(r{0.25em}){4-7}\cmidrule(r{0.25em}){8-11}
%\multicolumn{3}{c}{Dataset} & \multicolumn{2}{c}{Size of} & \multicolumn{2}{c}{Index}& \multicolumn{2}{c}{Size of} & \multicolumn{2}{c}{Index}\\
%\cmidrule(r{0.25em}){1-3}\cmidrule(r{0.25em}){4-5}\cmidrule(r{0.25em}){6-7}\cmidrule(r{0.25em}){8-9}\cmidrule(r{0.25em}){10-11}
%Euro-IX & PeeringDB & PCH & Intersection & Union & Jaccard & Overlap & Intersection & Union & Jaccard & Overlap\\
%\midrule
%\checked & \checked & \checked & 325 & 885 & 36.7\% & 69.0\% & 273 & 673 & 40.6\% & 73.0\%\\
%\checked & \checked & & 380 & 571 & 66.5\% & 80.7\% & 355 & 566 & 62.7\% & 80.5\%\\
%\checked & & \checked & 379 & 749 & 50.6\% & 80.5\% & 303 & 512 & 59.2\% & 81.0\%\\
%& \checked & \checked & 344 & 793 & 43.4\% & 71.7\% & 288 & 566 & 50.9\% & 77.0\%\\
%\bottomrule
%\end{tabular}
%\caption{Intersection and union of the IXP sets which are present in different combinations of
%datasets, as well as similarity indexes for the sets. The left columns show all IXPs, while the right columns only show
%those marked as active.
%\label{tab:matches-by-dataset}}
%\end{table*}

\begin{table}[t]
\centering
\scriptsize
\begin{tabular}{ccccrrrr}
\toprule
& & & \multicolumn{4}{c}{Active IXPs}\\
\cmidrule(r{0.25em}){4-7}
\multicolumn{3}{c}{Dataset} & \multicolumn{2}{c}{Size of} & \multicolumn{2}{c}{Index}\\
\cmidrule(r{0.25em}){1-3}\cmidrule(r{0.25em}){4-5}\cmidrule(r{0.25em}){6-7}
Euro-IX & PeeringDB & PCH & Intersection & Union & Jaccard & Overlap\\
\midrule
\checked & \checked & \checked & 273 & 673 & 40.6\% & 73.0\%\\
\checked & \checked & & 355 & 566 & 62.7\% & 80.5\%\\
\checked & & \checked & 303 & 512 & 59.2\% & 81.0\%\\
& \checked & \checked & 288 & 566 & 50.9\% & 77.0\%\\
\bottomrule
\end{tabular}
\caption{Intersection and union of the IXP sets which are present in different combinations of
datasets, as well as similarity indexes for the sets.
\label{tab:matches-by-dataset}}
\end{table}

We highlight that the datasets provide a lot of complementary information.
We interpret this, as well as the differences in IXP names, as indicators that
the datasets do not in general have a common source. We further
elaborate on this finding in the next section. In total, we find 441, 480 and 374 active
IXPs in the Euro-IX, PeeringDB and PCH datasets (after merging), respectively.
If we also consider inactive IXPs (\eg IXPs marked as
``defunct'' or ``unknown'')  there are 471, 480 and 657 IXPs in the Euro-IX,
PeeringDB and PCH datasets. Note that 43.1\perc of the IXPs present in
the PCH dataset are inactive. We make the compiled datasets  available
in~\cite{mappings}. Compared to the commonly-used PeeringDB, the combined
dataset includes information for 40.2\perc more active IXPs.

%If we also consider inactive IXPs (\eg IXPs marked as
%``defunct'' or ``unknown'')  there are 471, 480 and 657 IXPs in the Euro-IX,
%PeeringDB and PCH datasets in total (after merging), which become 441,
%480 and 374, respectively, if we only consider IXPs marked as \emph{active}.

%We note the high number of IXPs in the PCH dataset which do not match
%the others. This is partially due to the large number of ``defunct''
%and ``unknown'' IXPs present in the PCH dataset. Thus, we also
%provide our analysis when only considering IXPs marked as active
%(cf. Table~\ref{tab:matches-by-dataset}).

%A detailed, structured analaysis of the results
%follows in Sec~\ref{sec:analysis}; 

\section{Status, locations, and facilities}\label{sec:analysis}
%\section{Cross-comparing the datasets}\label{sec:analysis}
%%\section{Analysis}

%In this section we first compare the PeeringDB, Euro-IX, and PCH databases
%and then analyze the linked dataset that encompasses all three sources. Finally, 
%we summarize artifacts we have identified in the three datasets.

In this section we compare the PeeringDB, Euro-IX, and PCH databases
with respect to the geographical distribution of IXPs, the co-location facilities that house
IXPs, and the IXP status information. 

\subsection{Geographical distribution}\label{sec:analysis-geographical}

\begin{table}[t]
\centering
\scriptsize
\tabcolsep2.0pt
\begin{tabular}{lllrrr}
\toprule
\multicolumn{3}{c}{Location} & \multicolumn{3}{c}{Number of IXPs}\\
\cmidrule(r{0.25em}){1-3}\cmidrule(r{0.25em}){4-6}
Continent & Country & City & Euro-IX & PeeringDB & PCH\\
\midrule
Africa & \emph{Total} & & \emph{31} & \emph{25} & \emph{30}\\
\cmidrule(r{0.25em}){2-6}
\multirow{5}{*}{Asia Pacific} & \multirow{2}{*}{Japan} & Tokyo & 9 & 6 & 11\\
 & & \emph{Total} & \emph{17} & \emph{14} & \emph{23}\\
 \cmidrule(r{0.25em}){3-6}
& \multirow{2}{*}{Indonesia} & Jakarta & 4 & 8 & 9\\
& & \emph{Total} & \emph{6} & \emph{13} & \emph{16}\\
\cmidrule(r{0.25em}){3-6}
& \emph{Total} & & \emph{75} & \emph{88} & \emph{116}\\
\cmidrule(r{0.25em}){2-6}
Australia & \emph{Total} & & \emph{16} & \emph{20} & \emph{23}\\
\cmidrule(r{0.25em}){2-6}
\multirow{9}{*}{Europe} & Russian Federation & & 24 & 24 & 19\\
\cmidrule(r{0.25em}){3-6}
& \multirow{2}{*}{France} & Paris & 9 & 8 & 14\\
& & \emph{Total} & \emph{19} & \emph{20} & \emph{28}\\
\cmidrule(r{0.25em}){3-6}
& Germany & & 16 & 16 & 25\\
\cmidrule(r{0.25em}){3-6}
& \multirow{2}{*}{United Kingdom} & London & 7 & 12 & 10\\
& & \emph{Total} & \emph{15} & \emph{12} & \emph{22}\\
\cmidrule(r{0.25em}){3-6}
& Sweden & & 13 & 11 & 14\\
& Poland & & 11 & 12 & 10\\
& \emph{Total} & & \emph{201} & \emph{196} & \emph{200}\\
\cmidrule(r{0.25em}){2-6}
Middle East & \emph{Total} & & \emph{8} & \emph{8} & \emph{10}\\
\cmidrule(r{0.25em}){2-6}
\multirow{6}{*}{North America} & \multirow{4}{*}{United States of America} & New York & 8 & 7 & 14\\
 & & Los Angeles & 5 & 3 & 10\\
 & & Chicago & 4 & 4 & 9\\
 & & \emph{Total} & \emph{92} & \emph{89} & \emph{156}\\
 \cmidrule(r{0.25em}){3-6}
& Canada & & 13 & 16 & 17\\
& \emph{Total} & & \emph{110} & \emph{107} & \emph{179}\\
\cmidrule(r{0.25em}){2-6}
\multirow{2}{*}{South America} & Brazil & & 28 & 41 & 36\\
& \emph{Total} & & \emph{48} & \emph{55} & \emph{64}\\
\cmidrule(r{0.25em}){2-6}
\emph{World} & \emph{Total} & & \emph{490} & \emph{499} & \emph{687}\\
\bottomrule

\end{tabular}
\caption{IXPs in each database by continent. For each continent, we display the countries
and cities with the most IXPs. The values reported are based on raw data \emph{before} merging sibling
IXPs because some IXPs are distributed in multiple cities.
%IXPs in each database by continent. The ten top countries in each database are also
%displayed, as well as the largest cities. Note that the values reported are based on raw data \emph{before} merging IXP
%entities.
\label{tab:ixps-by-continent}}
\end{table}

All of the datasets contain information concerning the location of IXPs. Based on this, 
in ~\tref{tab:ixps-by-continent} we show the geographical distribution of the IXPs 
across the globe, and compare how different regions are represented in each dataset.
We observe that \emph{the geographical coverage of Euro-IX and PeeringDB is similar, while
PCH has somewhat richer coverage in terms of sheer IXP numbers} (including inactive IXPs).
On the continent level, Europe has the largest share of IXPs, which
corresponds to approximately 40\perc in the Euro-IX and
PeeringDB datasets and 30\perc in the PCH dataset. Interestingly, Euro-IX does not have substantially more IXPs represented in Europe than
the other datasets. The next largest region is North America, where PCH has much greater
numbers than the other datasets---as discussed in \xref{sec:analysis-status}, this is largely due to inactive IXPs.
PCH also has a greater number of IXPs for the Asia-Pacific region, with Euro-IX having the least. The other regions
are broadly similar. 
The ranking of the largest countries is also similar across the datasets. 
The largest cities differ more, with only major world cities being
consistently at the top of all of the datasets. 
In line with our expectations, it appears that more affluent regions have a better coverage
by IXPs.

\subsection{Facilities}\label{sec:analysis-facilities}
Euro-IX provides only the number of facilities for a limited subset of 106 IXPs (22\perc),
with these IXPs having a mean and median of 6 and 3 facilities, respectively. PCH
generally does not provide any facility-related information, although occasionally multiple
addresses are listed.
In contrast, PeeringDB contains detailed information about facilities, representing them
with separate database entities. There are 1,465 facilities listed, 365 of which are in the United States,
126 in Germany, 114 in the United Kingdom, 94 in France and 86 in the Netherlands.
The majority of the facilities \emph{are not} associated with an IXP, while 298 IXPs do
not report their facilities. 16 facilities are associated with neither IXP nor ISP entities.
These observations suggest that \emph{the information on the IXPs' facilities is limited}. 
Besides, 133 of the facilities associated with an IXP have more than one IXP present,
while 112 IXPs are present at more than one facility and 13 are present at more
than 10. This indicates that large IXPs are in reality geographically
distributed entities.%resembling ISPs. 
Understanding the drivers and implications
of this expansion and transformation that large IXPs undergo is an interesting subject for future work. 

%\bernhard{Not sure if it is smart to open another battle field with the 
%following side note.}
%\fontas{I moved this here, where it might fit better as an insight for future work}

\subsection{IXP status information}\label{sec:analysis-status}
The Euro-IX and PCH datasets contain information about the status of IXPs,
\ie whether or not they are currently active. Of all the IXPs in the Euro-IX dataset,
460 are marked as active, 23 as defunct and 7 as under construction. The PCH
dataset contains 392 marked active, 90 defunct, 43 planned, 6
deprecated, while 92 have an unknown status. In the PCH dataset 52 entries
have the status ``not an exchange''. Of the 379 common IXPs between these two datasets, 303
share an active status, while 9 share a defunct status. 10 of the matched entries appear
as defunct only in the PCH dataset and 4 only in the Euro-IX dataset. Overall, \emph{the status
information of the 379 linked IXPs is 82.8\perc consistent between the Euro-IX and PCH
datasets}.

PeeringDB contains no information on the status of IXPs. Still, a total
of 28 PeeringDB entries are marked as defunct in at least one of the Euro-IX (21
entries) or PCH (15 entries) datasets. It is noteworthy that of these 28 IXPs only
six report zero participants in PeeringDB, while the others usually report between
one and 20, with one IXP reporting 43 participants. We also checked the websites of 
IXPs marked as deprecated in Euro-IX or PCH,  but yet still reported
on PeeringDB. The results showed that most websites cannot be
reached or have extremely few members. For example, NWIX Missoula
reports only 4 active members, LIX (Luxembourg) has merged with
LU-CIX, and five websites don't report an active IXP any more.

%\todo{Minor: Did we check when these 28 IXPs were last updated?}

%The PeeringDB dataset contains no information on the status of IXPs.
%Still, of those entries matching the Euro-IX and PCH datasets, a total
%of 28 are marked as defunct in at least one of the Euro-IX (21 entries) or
%PCH (15 entries) datasets. It is noteworthy that of these 28 presumably
%defunct IXPs only six report zero participants in PeeringDB, while the
%others usually report between one and 20, with one presumably defunct
%IXP reporting 43 participants.
%\todo{Did we check when the presumably defunct IXPs were last updated}

%They appear to be roughly evenly distributed (15 in Europe, 11
%in North America, 7 in the Asia Pacific Region and 4 in Africa). 

Lastly, all but two of the IXPs appearing only in the Euro-IX dataset (38)
are marked as active. In contrast, half of the 259 IXPs which are only
present in the PCH dataset are either defunct (65) or have unknown
status (65), and only 56 of these IXPs are marked as active. Many of the
PCH-only IXPs are located in North America. Indeed, according to the
PCH dataset, \emph{North America has the largest number of defunct IXPs},
which is likely due to IXPs deployed in the early history of Internet
development.

%% (such as the well-known Metropolitan Area Exchange \emph{MAE-West}).
%% MAE-West is not defunct. Fontas

%\section{Complementarity and Accuracy}\label{sec:analysis-status}
\section{IXP participants}\label{sec:ixp-participants}

For many use cases, the participants (\ie peering ASes) of IXPs constitute the most 
important content of the datasets. Thus, we take a closer look at them in this section.  

%In this section, we ask the questions 1) how much additional information the
%linked dataset provides compared to using each database alone? and 2) how far
%the linked dataset is from possible "ground truths"? We highlight though that
%our main purpose is to analyze the similarities and differences of the three
%databases and that the linked dataset is a side-product of our comparison. 
%An in depth evaluation of its accuracy requires finding thorough "ground truth"
%data, which is beyond the scope of our preliminary analysis. 

\subsection{IXP-centric versus AS-centric view}\label{sec:analysis-membership}
%\subsubsection{IXP participants: the IXPs' point of view}\label{sec:analysis-membership}

\begin{figure}[t]
\centering
\subfloat[By IXP\label{fig:ccdf-membership-ixp}]{\includegraphics[width=0.49\columnwidth,clip=true,trim=0 0 0 0]{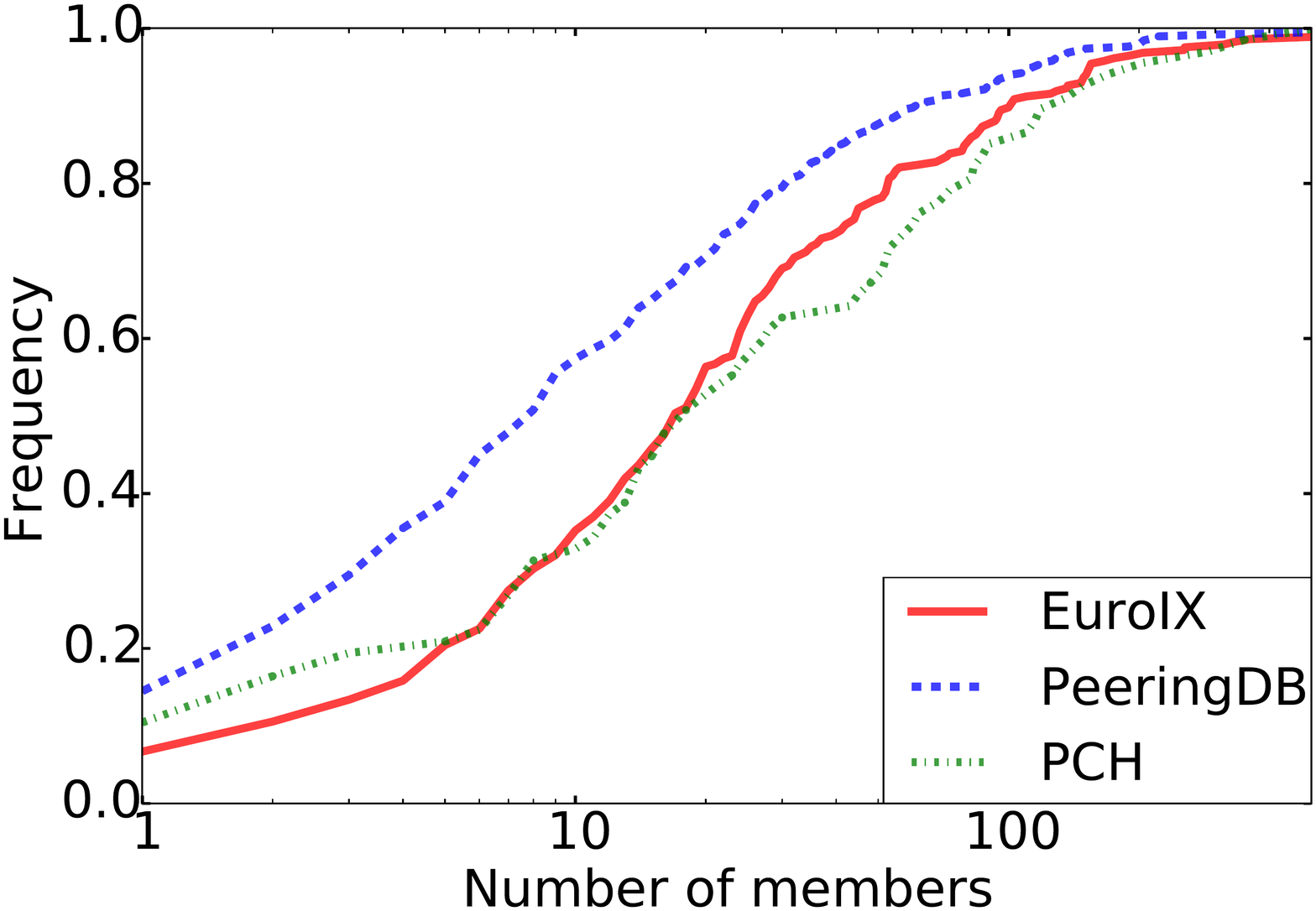}}
\subfloat[By ASN\label{fig:ccdf-membership-asn}]{\includegraphics[width=0.49\columnwidth,clip=true,trim=0 0 0 0]{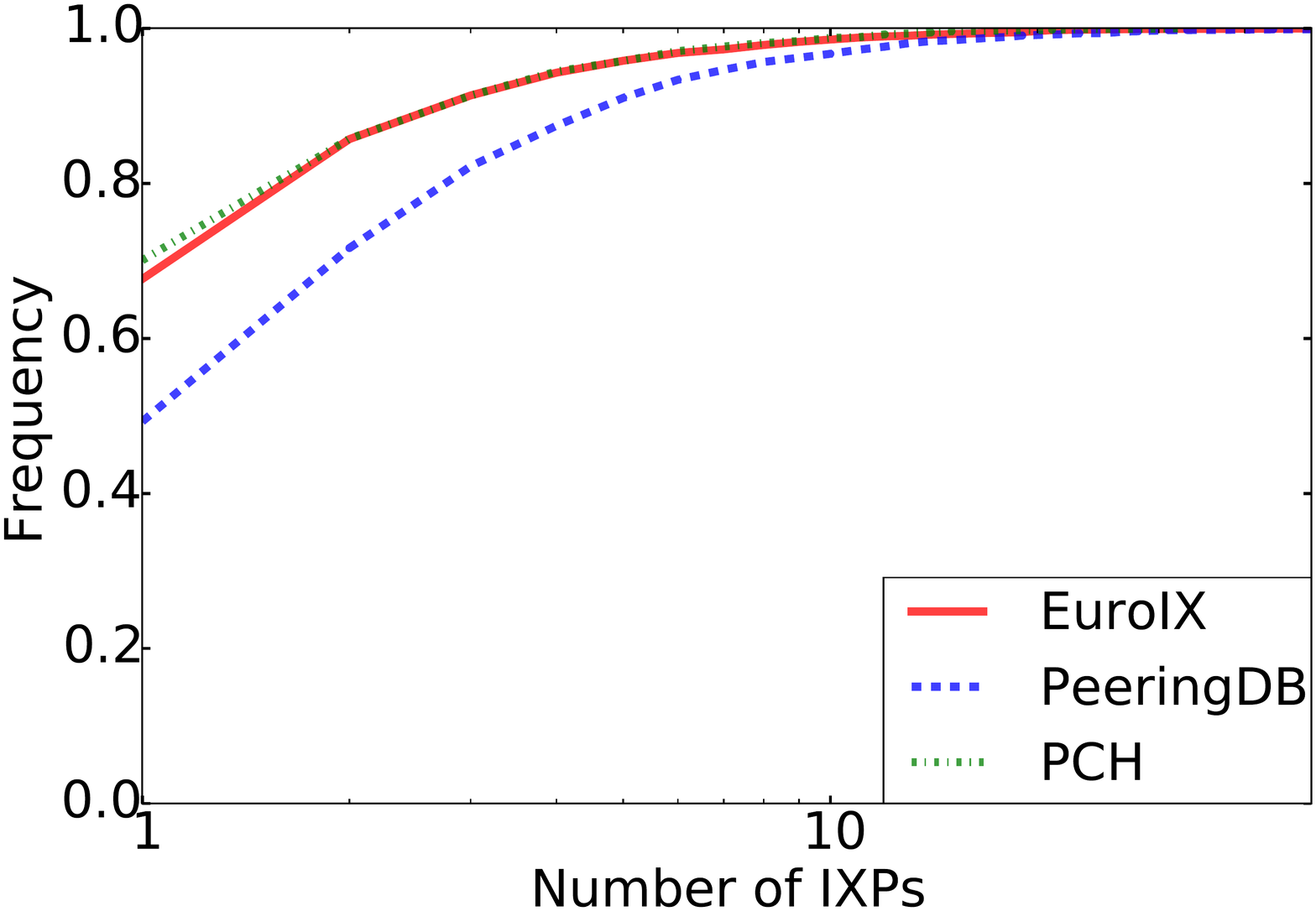}}
\caption{CDFs of the ASes per IXP (\fref{fig:ccdf-membership-ixp}) versus the IXPs per AS (\fref{fig:ccdf-membership-asn}), for each of the databases. In \fref{fig:ccdf-membership-ixp}, IXPs with no participants are omitted. 
%%In the case of PCH, a very substantial number of IXPs have no participants.
\label{fig:ccdf-membership}}
\end{figure}

Excluding IXPs which have no participants listed, the Euro-IX, PeeringDB and PCH
datasets have a mean of 44.3, 27.0 and 30.8 participants per IXP,
respectively, with corresponding medians of 17, 8 and 15. This suggests that PeeringDB
entries have on average considerably fewer IXP participants listed than Euro-IX entries.
\fref{fig:ccdf-membership-ixp} shows the distribution of participant counts for the three datasets.
We see that, in general, Euro-IX has the largest number of participants per IXP. 
Euro-IX provides an \emph{IXP-centric} view as its data is primarily self-reported by IXPs.
Besides, IXPs affiliated with Euro-IX typically have a high number of participants---a mean
of 104 and a median of 53, contrasting with a mean of 24 and a median of 13 for non-affiliates---as
a result of more complete reporting and also because many of the largest IXPs, \eg LINX, AMS-IX, and DE-CIX,
are Euro-IX affiliates. This indicates that large IXPs are generally better represented in the Euro-IX database.

On the other hand, 205 Euro-IX IXPs, 104 PeeringDB IXPs and 636 PCH
IXPs have no participants listed. 
%For PCH this means that 93\perc of all of the
%IXPs present do not have participant information. 
89\perc (53\perc) of the
Euro-IX (PCH) IXPs which have no participants listed are marked as active.
Interestingly enough, seven Euro-IX affiliate IXPs have no participants in the Euro-IX
database. Of these, only two separate IXPs appear in each one of the
other databases. One of these, CyrusOne, has a limited amount of
information about their IXP connectivity available in PeeringDB.  

%\todo{Vassilis: I cut out the following paragraph (commented out) because I agree with Bernhard: it does
%not add any important insight, and the Euro-IX bias towards large IXps has already been mentioned}
%\todo{Not sure what the take-away message of the next paragraph should
%  be. In my opinion, we can cut the whole paragraph. --Bernhard.}
%Euro-IX appears to have more participants listed for the largest IXPs,
%while PeeringDB has the values spread more evenly, and for PCH it is
%difficult to make a precise statement given the paucity of available
%data. The 85 IXPs present only in the PeeringDB dataset appear to also
%be geographically evenly distributed; the most are in the Asia-Pacific
%region.

%As mentioned in Sec.~\ref{sec:data_sources}, Euro-IX may have
%some bias towards larger IXPs.

%%\subsubsection{IXP participants: the participants' point of view}\label{sec:analysis-participants}

We further analyze IXP participants from the perspective of the
participating ASes. The Euro-IX dataset contains records of 6,697
ASes, connected to 1.9 IXPs on average. In PeeringDB, there are
3,784 ASes represented; these are connected to an average of 2.8 IXPs.
Finally, PCH contains 1,138 ASes, connected to an average of 1.4 IXPs. 
2,167 (Euro-IX), 1,999 (PeeringDB) and 201 (PCH)
ASes are connected to more than one IXP; 98, 127 and 5 are connected
to more than ten, respectively.
Table~\ref{tab:ixp-asn-membership} shows the ASNs which are connected
to the largest number of IXPs. We see that Packet Clearing House is
among the most prolific peers. PCH's ASN 3856 is used to acquire BGP
dumps, reflecting its strategy for data acquisition. PCH's ASN 42 is
used for hosting anycasted DNS zones. We also note the presence of large CDNs, like
Akamai. \fref{fig:ccdf-membership-asn} shows the distribution of participant counts from the
ASes' perspective for the three databases. The values of IXPs per AS for PeeringDB are
generally higher than the values for Euro-IX.  These differences likely stem from the
mechanisms with which the datasets are formed. 
In contrast to Euro-IX, PeeringDB provides an \emph{AS-centric} view as its
data is self-reported by ASes.

%\emph{The number of IXPs
%that an AS is connecting to has generally larger values in the PeeringDB dataset}.

\begin{table}[t]
\centering
\tabcolsep1.8pt
\scriptsize
\resizebox{\columnwidth}{!}{  
\begin{tabular}{llllrrr}
\toprule
& & & & \multicolumn{3}{c}{Number of IXPs}\\
\cmidrule(r{0.25em}){5-7}
ASN & Name & Policy & Network Type & Euro-IX & PeeringDB & PCH\\
\midrule
20940 & Akamai Technologies & Open & Content & 61 & 91 & 31\\
6939 & Hurricane Electric & Open & NSP & 66 & 84 & 32\\
15169 & Chief Telecom Inc. & Open & NSP & 60 & 76 & 24\\
3856 & Packet Clearing House & Open & Educ./Research & 50 & 74 & 21\\
42 & Packet Clearing House & Open & Educ./Research & 44 & 75 & 21\\
8075 & Microsoft & Selective & NSP & 37 & 59 & 22\\
22822 & Limelight Networks & Selective & Content & 41 & 39 & 18\\
15133 & EdgeCast Networks, Inc. & Open & Content & 25 & 31 & 18\\
16509 & Chief Telecom Inc. & Open & NSP & 21 & 44 & 7\\
10310 & Yahoo! & Selective & Content & 27 & 27 & 14\\
\bottomrule
\end{tabular}
}
\caption{The ASNs connecting to the largest number of IXPs (ranked by the sum). The ancillary
information is as reported by PeeringDB.\label{tab:ixp-asn-membership}}
\end{table}

\subsection{Complementarity of IXP participant data}\label{sec:analysis-similarity}

We build IXP-to-ASN \emph{links} for each dataset,
which represent $(\mathrm{IXP},\mathrm{ASN})$ memberships, and
perform set-theoretic operations on the extracted links using the Jaccard
and overlap indexes as introduced in \xref{sec:matching}. In
\tref{tab:by_continent_and_size} we compare the number and similarity
of the IXP participants by continent and IXP sizes.

\begin{table*}[t]
\centering
\scriptsize
\begin{tabular}{lrrrrrrrrr}
\toprule
 & \multicolumn{3}{c}{Number of links} & \multicolumn{2}{c}{Euro-IX/PeeringDB} & \multicolumn{2}{c}{Euro-IX/PCH} & \multicolumn{2}{c}{PeeringDB/PCH}\\
\cmidrule(r{0.25em}){2-4}\cmidrule(l{0.25em}r{0.25em}){5-6}\cmidrule(l{0.25em}r{0.25em}){7-8}\cmidrule(l{0.25em}){9-10}
Category & Euro-IX & PeeringDB & PCH & \quad Jaccard & Overlap & Jaccard & Overlap & Jaccard & Overlap\\
\midrule
\multicolumn{10}{l}{Continent}\\
\midrule
Africa & 247 & 163 & 27 & 23.5\% & 47.9\% & 2.24\% & 22.2\% & 9.83\% & 63.0\%\\
Asia Pacific & 1049 & 1105 & 516 & 28.4\% & 45.4\% & 22.3\% & 55.2\% & 22.1\% & 56.8\%\\
Australia & 353 & 470 & 49 & 20.7\% & 39.9\% & 6.07\% & 46.9\% & 8.81\% & 85.7\%\\
Europe & 7747 & 5370 & 1937 & 46.3\% & 77.3\% & 22.6\% & 92.0\% & 29.1\% & 85.1\%\\
Middle East & 41 & 32 & 27 & 40.4\% & 65.6\% & 47.8\% & 81.5\% & 63.9\% & 85.2\%\\
North America & 2059 & 2436 & 1009 & 35.1\% & 56.8\% & 25.9\% & 62.5\% & 27.2\% & 73.0\%\\
South America & 1088 & 693 & 2 & 38.0\% & 70.7\% & 0.0918\% & 50.0\% & 0.289\% & 100\%\\
\midrule
\multicolumn{10}{l}{Size of IXP}\\
\midrule
Less than 30 & 3375 & 3074 & 246 & 24.2\% & 40.8\% & 2.52\% & 36.2\% & 4.96\% & 63.8\%\\
30 to 59 & 1948 & 1324 & 277 & 31.5\% & 59.2\% & 11.1\% & 80.5\% & 11.4\% & 59.2\%\\
60 to 119 & 2837 & 2159 & 855 & 38.9\% & 64.8\% & 18.8\% & 68.3\% & 24.8\% & 69.9\%\\
120 to 239 & 2064 & 1749 & 1041 & 49.1\% & 71.8\% & 33.7\% & 75.1\% & 41.3\% & 78.3\%\\
240 or more & 2360 & 1963 & 1155 & 74.3\% & 93.9\% & 44.1\% & 93.2\% & 49.5\% & 89.4\%\\
\midrule
Total & 12584 & 10269 & 3574 & 40.1\% & 63.7\% & 20.5\% & 77.1\% & 25.0\% & 77.4\%\\
\bottomrule
\end{tabular}
\caption{The number of IXP-to-ASN links by category, and the Jaccard and overlap indexes between
each pair of datasets for each category. The categories used are \emph{continent} and \emph{IXP size}---the latter is
computed by averaging over all the datasets in order to yield a consistent classification scheme for the three datasets.\label{tab:by_continent_and_size}}
\end{table*}

%The Jaccard index of IXP-ASN links between Euro-IX and
%PeeringDB is at a mere 40\perc, which indicates that compared
%to PeeringDB, by simply combining with the Euro-IX dataset we could increase
%the available IXP membership information by 58.9\perc.  Overall, 
%the additional links in the union of all three datasets compared to
%PeeringDB is 66.3\perc.

The Jaccard index of IXP-ASN links between Euro-IX and
PeeringDB is at a mere 40\perc. Merging PeeringDB with
Euro-IX increases the available IXP membership information
by 58.9\perc.  This number goes to 66.3\perc when merging
PeeringDB both with Euro-IX and PCH. 
%Overall, the overlap index
%yields between 60\perc and 70\perc for each pair of datasets,
%indicating that the datasets are at least consistent to some degree.
Note that the similarity between the Euro-IX and PeeringDB participant
information is greatest in Europe, the region for which both datasets
have the largest quantity of membership information (links in \tref{tab:by_continent_and_size}).
In the case of Euro-IX, this constitutes well over half of all
participant information available. 75\perc
of the links in Europe (corresponding to 46\perc of all
links) are contributed by just the Euro-IX
affiliated IXPs. Other regions are reported more sparsely, yielding
lower similarity: North and South America have Jaccard indexes of
35\perc and 38\perc, respectively, and other regions have values under
30\perc. For the Middle East, the number of participants is so small
that the similarity is not meaningful.

As expected, the Jaccard index is much
lower for comparisons involving the PCH dataset due to the limited
membership data within the PCH dataset. In terms of the overlap
index, the PCH dataset has nearly the same (low) similarity to both
of the other datasets, but there are some notable differences between
regions: PCH is more in line with Euro-IX within Europe, and otherwise
closer to PeeringDB. However, these differences are small in regions with
a meaningful amount of information. 

Looking at the size categories in
\tref{tab:by_continent_and_size}, we find that larger IXPs have a
greater similarity, across all pairs of datasets. This holds for both
the Jaccard and overlap index. Unfortunately, PCH does not provide
participant information for the IXPs in the largest size category,
namely AMS-IX, DE-CIX (both Frankfurt and Hamburg), LINX, NIX.CZ
(Prague), PTT S\~{a}o Paulo, and SIX (Seattle). 

\section{Completeness of the IXP participant data}\label{sec:accuracy} 

In this section we do a first analysis of the accuracy of the IXP participant information extracted from the three databases. In particular, we
try to answer the question of the completeness of the collected information. We cross-compare the collected lists with IXP participant data extracted
from 1) live BGP sessions observed in IXP route collector BGP summary data; and 2) 40 IXP websites.

\subsection{Comparison with BGP data}
In~\xref{sec:matching} and~\xref{sec:analysis} we showed that by linking the available IXP
datasets we can significantly increase the available information about IXPs and their participants. 
In this section, we extract IXP participant information from BGP summaries collected by PCH
at 77 of their route collectors~\cite{pch-data} to compare and evaluate the completeness of the participant
information in all datasets, including the linked one. The BGP data include information about
established sessions with BGP peers over the IXP in contrast to the partially self-reporting 
origins of the other datasets. Thus, they are a ground truth for BGP peering sessions.
PCH tries to openly peer with all other IXP participants. Still, 
the data may miss participants who do not choose to peer with PCH. We assume that all peer
ASes seen by the IXP route collector peer over the IXP fabric. 
To verify this, we manually scanned
the next hop IPs and ASNs within the summary records to determine
which ASNs are actually peering at the IXPs by checking for IP addresses
from the prefixes assigned to the IXPs.
We used BGP data collected on the 19th of Sept 2014, \ie the same date 
as the other datasets, and again successfully linked the IXP identifiers of the 77
available PCH BGP route collectors with the IXP identifiers in the
other datasets using AS membership and IP address information. 
The route collectors contain location information in their name (typically an airport
code) which we utilized for further verification of the linked identifiers.

In~\tref{tab:accuracy} we report the number of IXP-to-ASN links by dataset
for the 77 IXPs with BGP route collectors and the Jaccard and overlap similarity
between the reference BGP data and the four other datasets. First, we find that
approximately 72\perc of the BGP IXP-to-ASN tuples are reported in the linked
dataset, while the corresponding figure is 65.8\perc for PeeringDB and lower
for the other datasets. 
Moreover, we find that Euro-IX and PeeringDB include many
IXP-to-ASN links which are not present in the BGP data. This indicates that the BGP
data is not complete, either. In particular, the route collectors report only approximately
56\perc of the membership contained in the databases. The underlying 
reasons include the fact that not all IXP participants may be willing to
peer with a route collector, and that the databases may contain stale data.

Besides, the validation dataset used in our study (and in all similar validation studies)
is subject to selection bias, \ie bias due to the IXPs and/or ISPs that provide useful 
information for validation. 
Indeed, looking at our set of 77 IXPs we find that the PeeringDB, PCH and
Euro-IX datasets are in larger agreement for this validation set than for the overall
comparison. \Eg PeeringDB and Euro-IX now have a Jaccard similarity of 53.1\perc as compared
to 40.1\perc in the earlier analysis (cf. \tref{tab:by_continent_and_size}).
We conclude that the figures presented on dataset completeness in the
77 IXPs may be positively biased.
This indicates that the information we have about the completeness of the
available IXP participant data, even after linking multiple databases, may
be still largely incomplete.

\begin{table*}[t]
\centering
\scriptsize
\begin{tabular}{rrrrrrrrrrrrr}
\toprule
\multicolumn{5}{c}{Number of links} & \multicolumn{2}{c}{BGP/UNION} & \multicolumn{2}{c}{BGP/Euro-IX} & \multicolumn{2}{c}{BGP/PeeringDB}  & \multicolumn{2}{c}{BGP/PCH}\\
\cmidrule(r{0.25em}){1-5}\cmidrule(l{0.25em}r{0.25em}){6-7}\cmidrule(l{0.25em}r{0.25em}){8-9}\cmidrule(l{0.25em}r{0.25em}){10-11}\cmidrule(l{0.25em}){12-13}
BGP & UNION & Euro-IX & PeeringDB & PCH & \quad Jaccard & Overlap & Jaccard & Overlap & Jaccard & Overlap & Jaccard & Overlap\\
\midrule
6,425 & 8,121  & 6,087 & 5,749 & 3,547 & 46.1\% & 71.5\% & 42.2\% & 61.0\% & 45.1\% & 65.8\% & 35.3\% & 73.4\% \\
\bottomrule
\end{tabular}
\caption{The number of IXP-to-ASN links by dataset for the 77 IXPs with BGP route collectors; and the Jaccard and overlap indexes between each dataset and the ground truth links extracted from the BGP route collectors. UNION denotes the linked dataset containing PeeringDB, Euro-IX, and PCH.\label{tab:accuracy}}
\end{table*}

%\subsection{How can the data be used}
%\todo{optional}

\subsection{Comparison with IXP website data}

%Besides evaluating the completeness of the datasets, we also checked
%IXP websites to resolve inconsistencies between the three datasets
%about deprecated IXPs. In particular, we checked the websites of 26
%IXPs marked as deprecated in Euro-IX or PCH,  but yet still reported
%on PeeringDB. The results showed that most websites cannot be
%reached or have extremely few members. For example, NWIX Missoula
%reports only 4 active members, LIX (Luxembourg) has merged with
%LU-CIX, and five websites don't report an active IXP any more.

We extracted participant lists from IXPs' websites as an additional 
source of cross-verification. In particular, we designed custom crawlers for
40 IXP websites in total, which include \first the 20 largest IXPs by number of 
participants, and \second 20 randomly selected IXPs. We selected two sets of IXPs
to mitigate the problem of the selection bias we discussed above. 
%based on the union of
%PeeringDB, Euro-IX, and PCH. 
If the website of an IXP did not list participant information, then we selected a further IXP either
by size or randomly from the two lists above. The website data were collected
during the 2nd half of August 2015. At the same time we extracted and linked fresh
data from Euro-IX, PeeringDB, and PCH for the selected IXPs to compare fairly with website data. 

\begin{figure}
\centering
%\subfloat[By IXP\label{fig:ccdf-membership-ixp}]{\includegraphics[width=0.49\columnwidth,clip=true,trim=0 0 0 0]{plots/cdf-ixp-membership-norm}}
%\subfloat[By ASN\label{fig:ccdf-membership-asn}]{\includegraphics[width=0.49\columnwidth,clip=true,trim=0 0 0 0]{plots/cdf-asn-memberships}}
\includegraphics[width=\columnwidth,clip=true,trim=0 0 0 0]{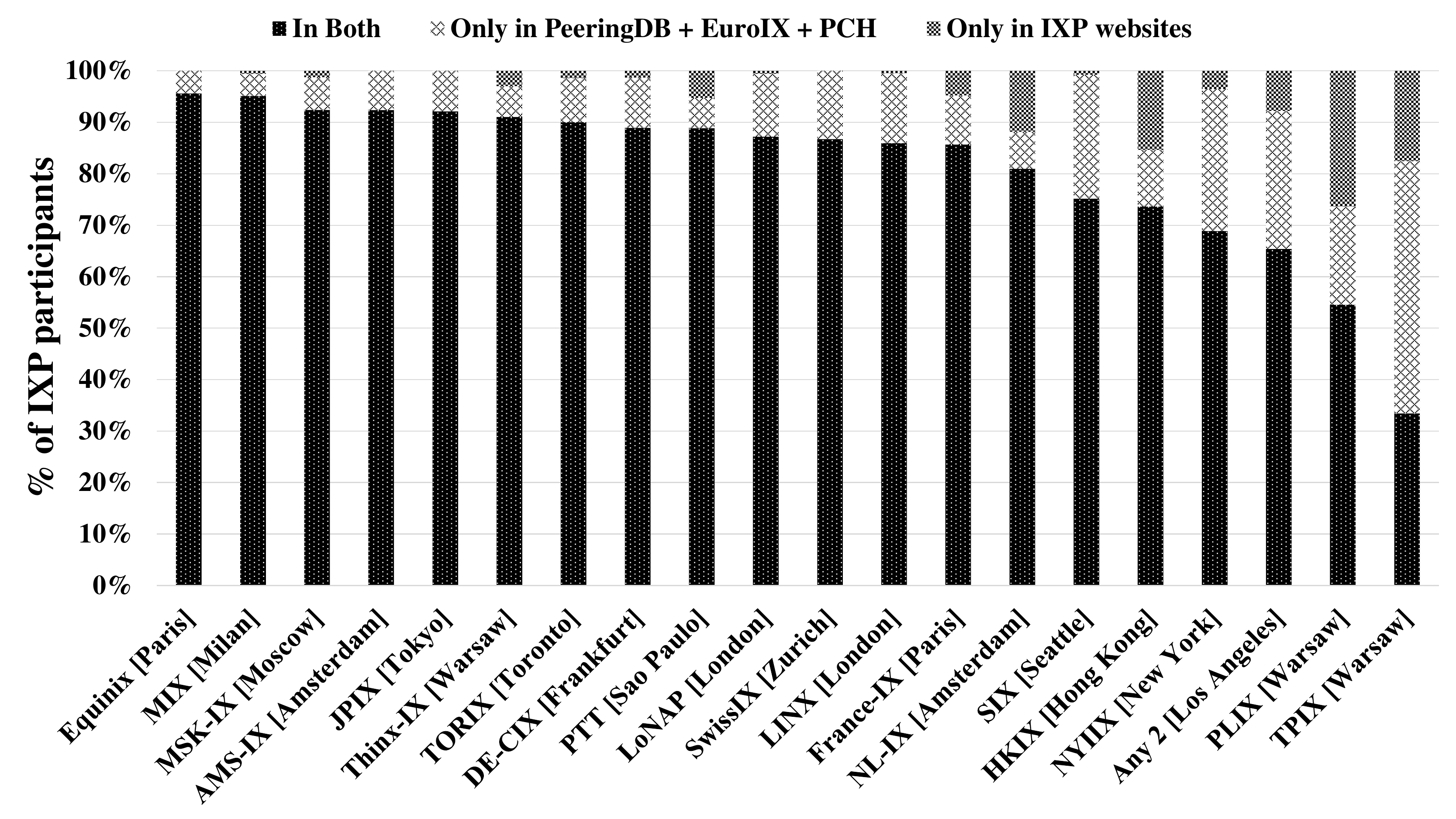}
\caption{Common and complementary participant information in IXP websites and in the union of PeeringDB, Euro-IX, and PCH datasets.
We show the top-20 IXPs with public participant data in their websites.\label{fig:top-20-websites}}
\end{figure}

From IXPs' websites, we extracted in total 6,182 IXP-to-ASN \emph{links} for the top-20 IXPs and 1,181
\emph{links} for the 20 random IXPs. 
We find that 94\perc of the links in the top-20 IXPs are reported in the union of
PeeringDB, Euro-IX, and PCH. This number changes to 85\perc for the
20 random IXPs. 
In \fref{fig:top-20-websites} we show the common information (\ie the Jaccard index)
between the websites and the linked dataset, and the information only in one of the
two sources for each of the top-20 IXPs. We order IXPs by the percentage of common
links. We see that for most websites the fraction of common links is above 80\perc. 
For many IXPs, we observe that the linked datasets contain more IXP-to-ASN links
than the websites of the IXPs. Only 6\% of the links are present only on websites.
In contrast, 14\perc of the links are present only in the linked dataset. Interestingly, 
this shows that the union of the three databases contains more information about 
IXP participants than the websites of the IXPs themselves.

\balance
\section{Conclusions and Future Work}\label{sec:conclusions} 

%We would
%like to point out each of the dataset collectors performs an
%outstanding job given such a herculean task, often lacking the direct
%support from IXPs and participant ASes. 

The quest for representative datasets is perpetual
for the research community. Taking into account the rising interest
in IXP-related data, in this work we \first compared three rich IXP datasets in order
to assess their strengths and weaknesses, and \second combined them 
in order to improve the completeness of the publicly available IXP
data. Our results show that the three datasets have similar geographical
coverage, with PCH having many more IXPs, but also many inactive
ones. In addition, PeeringDB has an AS-centric bias, while Euro-IX
has an IXP-centric bias due the nature of the self-reporting methodologies
used by the two providers. PCH includes very little information 
about IXP members. Furthermore, our results show that the datasets
have partially common as well as rich complementary information. With
respect to complementary, we show for example that by linking the datasets
we increase the number of IXP records by 40.2\perc compared to using solely % This doesn't agree with the 50.9% figure used elsewhere!
PeeringDB. Even more complementary information is available
for IXP member information, which previous studies have also shown to be
incomplete in PeeringDB~\cite{peeringDB-accuracy,peeringdb-routing-ecosystem}.
Finally, to aid future research, we have made the dataset snapshots as well as the mappings we constructed available
to the public, together with the code used to construct them~\cite{mappings}.

%Some of the datasets
%offer better quantity for certain geographical regions, \eg Euro-IX
%for Europe and PeeringDB for the US. On the other hand, the similarity
%between the datasets w.r.t. the IXP participants is surprisingly low, which
%indicates high degree of complementary information,  

%So far we have established that datasets on IXPs are, by and large,
%accurate, yet not complete, and that combining the datasets can help
%towards higher completeness. Still, comparing with BGP data shows that
%at least 25\perc of the membership information is missing from the
%combined dataset. 

Still, our results show that while the datasets are partially
consistent, they are also incomplete. In particular, the datasets
appear to be largely in agreement on the existence of IXPs, and certain
attributes such as their operational status.  Some of the datasets
offer better quantity for certain geographical regions, \eg Euro-IX
for Europe and PeeringDB for the US.  However the consistency between
the datasets w.r.t. the IXP participants is surprisingly low. We have
to stress that it is unclear to which degree these differences stem
from under-reporting, resp., from over-reporting such as out-aged
information.  Our study is a first step towards an in-depth analysis of IXP
datasets. The study opens a number of questions for future work.
We would like to understand how the datasets can be
cleverly combined, exploiting their individual strengths to improve
the accuracy of the available data. In particular, the ground truth behind the available IXP data
is still elusive and hard to determine. Other sources of possible
ground truth we did not explore in this work are: \first 
IXPs' looking glass servers, \second IXPs' newsletters, and \third event/feeds
at IXP websites, which announce new IXP members. 
A final line of enquiry is understanding the growth trends and
consistency of the IXP datasets over time within the evolving Internet peering
ecosystem.

\section*{Acknowledgments}\label{sec:acknowledgments}
We want to thank Euro-IX, PeeringDB and Packet Clearing House for providing free, publicly available sources
of information on Internet Exchange Points. In particular, we want to thank the staff of Euro-IX and
Packet Clearing House for providing us information about how data is collected for those datasets.
This work has been partly funded by the European Research Council Grant Agreement no. 338402.

\bibliographystyle{acm}
%\scriptsize
\small
\bibliography{refs}
\end{document}